\documentclass[debug,overfull]{epl}

\usepackage{graphicx}
\usepackage{dcolumn}
\usepackage{bm}
\usepackage{psfrag}
\usepackage{epsfig}
\usepackage{amsmath}
\usepackage{amssymb}
\usepackage{color}
\usepackage{fancyheadings}
\usepackage{mathbbol}

\newcommand{\nn}{\nonumber}
\newcommand{\bra}{\langle}
\newcommand{\ve}{\vert}
\newcommand{\ket}{\rangle}

\title{Scaling behavior of interactions in a modular quantum system and the existence of
local temperature}
\shorttitle{Scaling behavior of \dots}
\author{M. Hartmann\inst{1,2} \and J. Gemmer \inst{2} \and G. Mahler \inst{2} \and O. Hess \inst{3}}
\institute{
  \inst{1} Institute of Thechnical Physics, DLR Stuttgart -  Pfaffenwaldring 38-40, D-70569 Stuttgart,
Germany\\
  \inst{2} Institute of Theoretical Physics I, University of Stuttgart - Pfaffenwaldring 57,
D-70550 Stuttgart, Germany\\
  \inst{3} Advanced Technology Institute, University of Surrey -  Guilford, Surrey,  GU2 7XH, UK
}
\pacs{05.30.-d}{First pacs description}
\pacs{05.70.Ce}{Second pacs description}
\pacs{65.80.+n}{Third pacs description}

\begin{document}

\maketitle

\begin{abstract}
We consider a quantum system of fixed size consisting of a regular chain of $n$-level subsystems,
where $n$ is finite.
Forming groups of $N$ subsystems each, we show that the strength of interaction between the
groups scales with $N^{- \, 1/2}$. As a consequence, if the total system is in a thermal state with
inverse temperature $\beta$, a sufficient condition for subgroups of size $N$ to be approximately
in a thermal state with the same temperature is $\sqrt{N} \gg \beta \: \overline{\delta E}$,
where $\overline{\delta E}$ is the width of the occupied level spectrum of the total system.
These scaling properties indicate on what scale
local temperatures may be meaningfully defined as intensive variables.
This question is particularly relevant for non-equilibrium
scenarios such as heat conduction etc. 
\end{abstract}

\section{Introduction}

Thermodynamics has successfully been applied to the description of macroscopic systems since more than a
century.
For that reason this theory enjoys widespread  acceptance.
Nevertheless its microscopic foundation is, in most cases, not well understood.

The theory of thermodynamics is based on the notions of extensive and
conjugate intensive thermodynamic variables.
The existence of the thermodynamical limit in a mathematical sense
has been shown for many fundamental cases \cite{Fisher1964,Ruelle1969,Lebowitz1969}.
The standard proofs are based on the idea that, as the spatial extension increases,
the surface of a region in space grows slower than its volume.
If the interaction potential is short-ranged enough, one can show that the ''effective'' interactions
between one region and another become negligible in the limit of infinite size, implying extensivity.

However, the scaling behavior of the interactions with the size of those
regions has, to our knowledge, not been studied in any detail yet. While it has been shown that they vanish
in the thermodynamical limit, it remains unclear what happens in situations that are in some sense only
close to this limit. There is not even a precise understanding
of what ''close'' means in that case. 

For standard applications of thermodynamics this might not pose any serious problem since the
number of particles within any region is so large that deviations from infinite systems may
safely be neglected.
Nevertheless the differences should become important when the considered regions are significantly smaller.
It is here where a quantum approach becomes imperative \cite{GemmerOtte2001}.

The applicability of thermodynamical concepts to mesoscopic or even microscopic
systems has intensively been discussed
in recent years \cite{Jensen1985}, since nano-scale physics has attracted increased attention
\cite{Cahill2003}.
Due to the advance of experimental techniques the measurement of thermodynamic
quantities like temperature with a spatial resolution on the nanometer scale seems within reach
\cite{Gao2002,Pothier1997,Aumentado2001,Cahill2003}.
These techniques have already been applied for a new type of scanning microscopy,
using a temperature sensor \cite{Williams1986,Varesi1998}, that shows resolutions below 100nm.
An important
question thus arises \cite{Cahill2003}: Down to what spatial scale does a
meaningful notion of temperature exist at all? 

In this paper we consider a quantum system of fixed Hilbert-space dimension $dim$,
composed of identical elementary $n$-level subsystems. We form
$n_G$ identical groups of $N$ subsystems each ($dim = n^{n_G N}$) and
show that the interaction strength between the groups
scales inversely proportional to $\sqrt{N}$.
Based on this scaling relation and assuming the total system to be in a thermal state
we analyse, for what group size $N$ a thermodynamical description of the individual group
is appropriate.
The results are confirmed by numerical studies for a chain of 8 spins.
For this chain partitions into two 4-spin-groups, four 2-spin-groups and eight single spins are
considered.
%

\section{Scaling Law}

We consider a chain of identical $n$-level systems with identical nearest neighbour interactions.
The Hamiltonian of such a linear chain may be written as
\begin{equation}\label{hamil}
H = \sum_{j} h_{loc}(j) + h_{int}(j,j+1),
\end{equation}
where the index $j$ labels the elementary subsystems. The local terms $ h_{loc}(j)$ and the
nearest neighbour interactions $h_{int}(j,j+1)$ have the form \cite{Mahler1998}:
\begin{eqnarray}
\label{coeff}
h_{loc}(j) & = & \frac{n}{2} \: \sum_{\alpha} \: A_{\alpha} \: \sigma_{\alpha}(j) \nn \\
h_{int}(j,j+1) & = & \frac{n^2}{4} \:
\sum_{\alpha, \beta} \: C_{\alpha \beta} \: \sigma_{\alpha}(j) \: \sigma_{\beta}(j+1)
\end{eqnarray}
Here, the $\sigma_{\alpha}(j)$ are $SU(n)$ generators with
$\alpha, \beta = 1, 2, \dots, n^2 -1$, $n$ being the dimension of one subsystem.
For the $SU(n)$ generators we adopt the trace relations \cite{Mahler1998}:
\begin{eqnarray}
\label{trace}
\textrm{Tr} \left[ \sigma_{\alpha}(i) \right] & = & 0 \nn \\
\textrm{Tr} \left[ \sigma_{\alpha}(i) \: \sigma_{\beta}(j) \right] & = &
2 \: \frac{dim}{n} \: \delta_{i j} \: \delta_{\alpha \beta}
\end{eqnarray}
where the trace (Tr) has to be taken over the whole system of dimension $dim$.
As a consequence the trace over one elementary subsystem reads:
$\textrm{Tr}_{j} \left[ \sigma_{\alpha}(j) \: \sigma_{\beta}(j) \right] = 2 \: \delta_{\alpha \beta}$.
The coefficients in equation (\ref{coeff}) are then given by
$A_{\alpha} = \textrm{Tr}(H \, \sigma_{\alpha}(i)) \, / \, dim$ and 
$C_{\alpha \beta} = \textrm{Tr}(H \, \sigma_{\alpha}(i) \, \sigma_{\beta}(i+1)) \, / \, dim$,
respectively, and taken to be independent of $i$.
We assume periodic boundary conditions and $H$ to be traceless.

If we now form $n_G$ groups of $N$ subsystems each, we can split the Hamiltonian $H$ into two parts,
\begin{equation}
\label{hsep}
H = H_N^{(0)} + I_N
\end{equation}
where $H_N^{(0)}$ is the sum of the Hamiltonians of the individual groups and
$I_N$ describes the interaction between each group and its nearest neighbour,
\begin{equation}
I_N = \sum_i h_{int}(i N, i N + 1)
\end{equation}
We label the eigenstates of the total Hamiltonian $H$
and its energies by greek indices and eigenstates and energies
of the group Hamiltonian  $H_N^{(0)}$ by latin indices, i. e.
\begin{equation}
\label{prodstate}
H \: \ve \mu \ket = E_{\mu} \: \ve \mu \ket \enspace \enspace \textrm{and} \enspace \enspace
H_N^{(0)} \: \ve j_N \ket = E_{j_N} \: \ve j_N \ket
\end{equation}
Obviously, the $\ve j_N \ket$ are simply products of group eigenstates. We now proceed to compare
two characteristic quantities of the system according to equation (\ref{hsep}).

First, consider the identity
\begin{equation}
\label{condvar}
\sum_{\mu} \left( E_{\mu} - E_{j_N} \right)^2 \: \ve \bra j_N \ve \mu \ket \ve^2 =
\bra j_N \ve I_N^2 \ve j_N \ket
\end{equation}
where the left hand side can be interpreted as
the mean squared energy distance between a level $E_{\mu}$
and the energy $E_{j_N}$.
The average of equation (\ref{condvar}) over all $j_N$ is identical with
$\overline{\textrm{I}}_N^2$, where
\begin{equation}
\label{avI}
\overline{\textrm{I}}_N \equiv \sqrt{\frac{\textrm{Tr} \left( I_N^2 \right)}{dim}} =
\frac{n}{2}
\sqrt{n_G \sum_{\alpha, \beta} C_{\alpha \beta}^2}
\end{equation}
Here we have used equation (\ref{trace}).

The second quantity of interest is the width of the distribution of the energy levels of the
total system $\overline{\delta E}$ around the mean energy
$\overline{E} = \textrm{Tr}(H) / dim = 0$:
\begin{equation}
\overline{\delta E} \equiv \sqrt{\sum_{\nu} \frac{\left( E_{\nu} - \overline{E} \right)^2}{dim} } =
\sqrt{\frac{\textrm{Tr} \left[ \left( H - \overline{E} \right)^2 \right]}{dim}}
= \sqrt{N} \, \frac{n}{2}
\sqrt{n_G \left( \sum_{\alpha, \beta} C_{\alpha \beta}^2 + \frac{2}{n}
\sum_{\alpha} A_{\alpha}^2 \right)}\label{levelvar}
\end{equation}
Here we have again used equation (\ref{trace}).
Combining equations (\ref{avI}) and (\ref{levelvar}), we get the following scaling law:
\begin{equation}
\label{scalelaw}
\overline{\textrm{I}}_N \le \frac{1}{\sqrt{N}} \: \overline{\delta E}
\end{equation}
Note that this law is a property of the Hamiltonian (\ref{hamil}) and does not
depend on the state of the system.
The equality sign applies, if the local terms vanish, $A_{\alpha} = 0$. Equation (\ref{scalelaw})
does not depend on the convention used in equation (\ref{trace}), since normalisations cancel.
It is straightforward to see that the same scaling law holds for a three dimensional lattice with cubes
of $N^3$ subsystems as the subgroups.

Relation (\ref{scalelaw}) has been derived here for a perfectly homogenous system.
However, it is evident by means of stationary perturbation theory that
it still holds approximately for systems with small disorder.
%

\section{Distributions}

We now use the scaling law (\ref{scalelaw}) to estimate the
density matrix elements of the groups of $N$ subsystems assuming
that the total system is in a thermal state with the density matrix
\begin{equation}
\label{candens}
\bra \mu \ve \hat \rho \ve \nu \ket = \frac{e^{- \beta E_{\mu}}}{Z} \: \delta_{\mu \nu}
\end{equation}
in the eigenbasis of $H$. Here, $Z$ is the partition sum and $\beta = (k_B T)^{-1}$
the inverse temperature.
Transforming the density matrix (\ref{candens}) into the eigenbasis of $H_N^{(0)}$ we obtain
\begin{equation}
\label{newrho}
\bra j \ve \hat \rho \ve j \ket \: = \:
\sum_{\mu} \: \ve \bra j \ve \mu \ket \ve^2 \: \, \frac{e^{- \beta E_{\mu}}}{Z} \: = \:
\frac{e^{- \beta E_j}}{Z} \: \sum_{\mu} \: \ve \bra j \ve \mu \ket \ve^2 \:
\exp \left(- \beta ( E_{\mu} - E_j ) \right)
\end{equation}
(For simplicity we skip the index $N$ from now on).

A thermodynamical description of the groups of $N$ subsystems by canonical density matrices with
the same inverse temperature $\beta$ is appropriate, if
$\bra j \ve \hat \rho \ve j \ket$ was approximately proportional to $\exp ( - \beta E_j )$.
Truncating the sum to terms with $\beta \ve E_{\mu} - E_{j} \ve \ll 1$ (which will be motivated below),
we can expand the
second exponential in the rhs of equation (\ref{newrho}) into a Taylor series around $E_j$
up to second order,
\begin{equation}
\label{newrho2}
\bra j \ve \hat \rho \ve j \ket \approx
\frac{e^{- \beta E_j}}{Z} \, \left(1 - \beta \bra j \ve I \ve j \ket +
\frac{\beta^2}{2} \bra j \ve I^2 \ve j \ket \right),
\end{equation}
where we have used
$\sum_{\mu} E_{\mu} \: \ve \bra j \ve \mu \ket \ve^2=E_j \: + \: \bra j \ve I \ve j \ket$ 
and equation (\ref{condvar}). The second order term has to be taken into account, since
the sum over all energies $E_{\mu}$ adds up positive and negative contributions for the first order terms
but only positive ones for the second order terms. Since
$\bra j \ve I \ve j \ket^2 \leq \bra j \ve I^2 \ve j \ket$,
both correction terms are small if
\begin{equation}
\label{condvar2}
\beta \sqrt{\bra j \ve I^2 \ve j \ket} \ll 1.
\end{equation}

Equation (\ref{condvar2}) justifies the above truncation
if the distribution 
$w_j(\mu) = \ve \bra j \ve \mu \ket \ve^2$ times the density of states $\eta ( E_{\mu} - E_{j} )$
decayed faster than
$\exp (- \beta \ve  E_{\mu} - E_{j} \ve)$ for $\ve E_{\mu} - E_{j} \ve > \sqrt{\bra j \ve I^2 \ve j \ket}$
and fixed $j$. We numerically verified this behavior for a class of systems as shown in figure
\ref{condprobplot}.
If, on the other hand, the truncation of the sum was not possible, the rhs of equation (\ref{newrho2}) would
contain additional correction terms invalidating a local thermodynamical description.

In the basis $\ve j \ket$, the off-diagonal elements of the density matrix
$\sum_{\mu} \bra j \ve \mu \ket \bra \mu \ve j' \ket \exp (- \beta E_{\mu})$ vanish
for $\ve E_{j} - E_{j'} \ve > 2 \overline{\textrm{I}}$ because
$\ve \bra j \ve \mu \ket \bra \mu \ve j' \ket \ve \approx 0$.
When $\ve E_{j} - E_{j'} \ve < 2 \overline{\textrm{I}}$, one can use the same approximation as for
the diagonal terms, where now the zero order term is zero. In the first and second order
corrections each term of the sum
carries a phase and thus these corrections are smaller than for diagonal elements.

Combining equations (\ref{scalelaw}) and (\ref{condvar2}) we thus conclude
that the condition
\begin{equation}
\label{scalecond}
\sqrt{N} \gg \beta \: \overline{\delta E}.
\end{equation}
is sufficient to allow for an approximate local thermodynamical description for a group size $N$.
In equation (\ref{scalecond}) we have used that 
$\overline{\textrm{I}}^2$ is the arithmetic mean of all $\bra j \ve I^2 \ve j \ket$: these
are positive quantities and therefore
$\beta \overline{\textrm{I}} \ll 1$ implies $\beta \sqrt{\bra j \ve I^2 \ve j \ket} \ll 1$ for almost
all states $\ve j \ket$. Equation (\ref{scalecond}) is the main result of our paper.

%

\section{Numerical results}

To test the condition (\ref{scalecond}) and the approximations involved,
we investigate a chain of 8 spins with a Hamiltonian of the form (\ref{hamil}) rewritten as
\begin{equation}
H = \frac{\Delta E}{2} \sum_{j=1}^8 \sigma_z (j) + \lambda \sum_{j=1}^8 \sum_{\alpha,\beta=1}^3
c_{\alpha \beta} \: \sigma_{\alpha} (j) \: \sigma_{\beta} (j+1)
\end{equation}
Without loss of generality, we restrict the local part to terms in $\sigma_z$ only; different local
terms would merely imply a rotation of the coordinate system. Periodic boundary conditions are chosen,
$\sigma_{\alpha}(9) = \sigma_{\alpha}(1)$. $\lambda$ is a scaling factor for the interactions and
the $\sigma_{\alpha}$ ($\alpha = 1,2,3$) are the Pauli matrizes. In the following all energies
(i.e. $\lambda$, $\beta^{-1}$) will be taken to be in terms of $\Delta E$.

For each realisation the matrix elements $c_{\alpha \beta}$ are randomly chosen from the interval $[-1,1]$
with equal weight.
Based on 100 such realisations, we find the average of $\overline{\delta E}$ to be $\approx 5 \lambda$.
Therefore, from equation (\ref{scalecond}), a local thermodynamical description is expected to be
appropriate for $\sqrt{N} \, / \, (\beta \lambda) \gg 5$.

We consider three different partitions of this system into groups of adjoining subsystems
as described in  equation (\ref{hsep}).
The partitions are: two 4-spin-chains, four 2-spin-chains and eight single spins.
For the inverse temperature of the total system $\beta$ and the interaction strength $\lambda$
we consider the values $\beta \lambda = 0.1, 0.2, 0.3, 0.4$ which implies
$2.5 \le \sqrt{N} \, / \, (\beta \lambda) \le 20$.

\begin{figure}[h!]
\psfrag{1.1}{\small \hspace{-0.2cm} 1.0}
\psfrag{0.8}{\small \hspace{-0.2cm} 0.8}
\psfrag{0.6}{\small \hspace{-0.2cm} 0.6}
\psfrag{0.4}{\small \hspace{-0.2cm} 0.4}
\psfrag{0.2}{\small \hspace{-0.2cm} 0.2}
\psfrag{-4}{\small \raisebox{-0.1cm}{-4}}
\psfrag{-2}{\small \raisebox{-0.1cm}{-2}}
\psfrag{0}{\small \raisebox{-0.1cm}{0}}
\psfrag{1}{\small \raisebox{-0.1cm}{1}}
\psfrag{2}{\small \raisebox{-0.1cm}{2}}
\psfrag{3}{\small \raisebox{-0.1cm}{3}}
\psfrag{4}{\small \raisebox{-0.1cm}{4}}
\psfrag{r}{$\overline{\textrm{I}} \, / \, \overline{\delta E}$}
\psfrag{n}{$N$}
\psfrag{En}{$x$}
\psfrag{p}{$\ve \bra \mu \ve j \ket \ve^2 \, \eta(E_{\mu}-E_j)$}
\twofigures[scale=0.65]{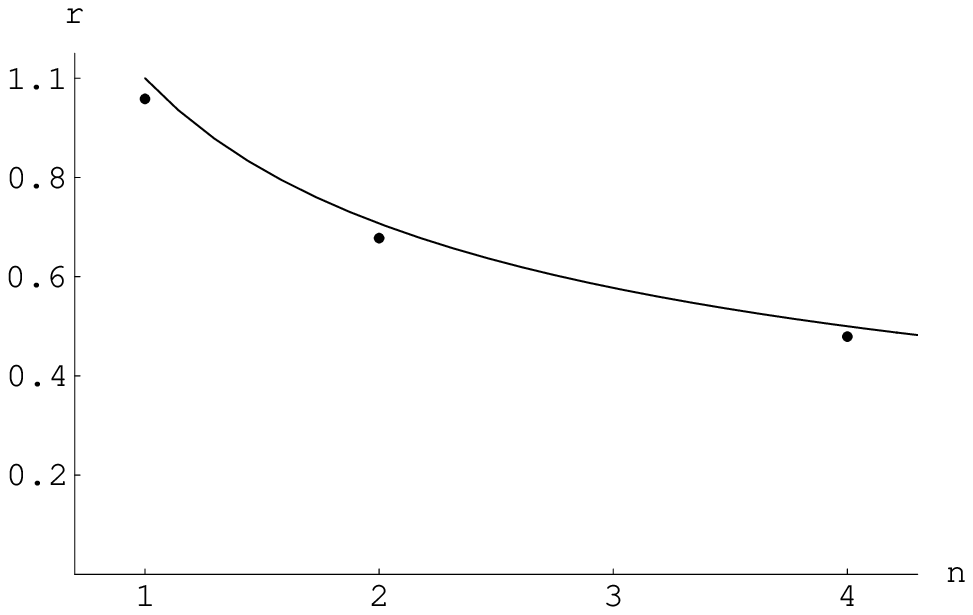}{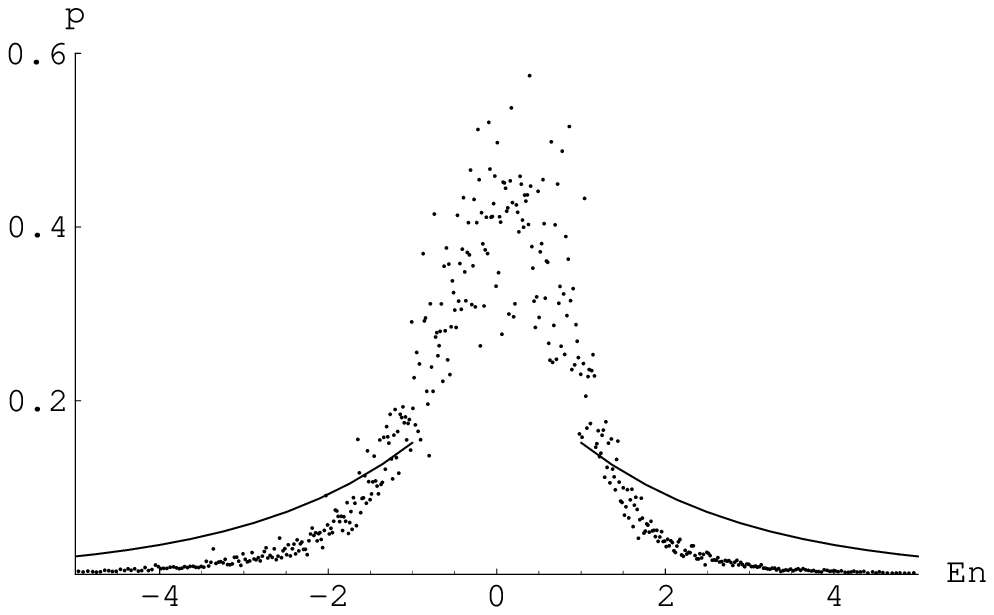}
\caption{The ratio of the average interaction strength and the level spreading
$\overline{\textrm{I}} / \overline{\delta E}$ (dots) as a function of the number
of subsystems per group $N$ for $\lambda = \Delta E$. The line shows $1 / \sqrt{N}$.}
\label{int}
\caption{The conditional probabilities $\ve \bra \mu \ve j \ket \ve^2$ weighted with the density of
states $\eta(E_{\mu})$ as a function of
$x = (E_{\mu} - E_j) / \sqrt{\bra j \ve I^2 \ve j \ket}$. The lines $\, 0.25 \, \exp (- 0.5 \, \ve x \ve)$
are shown for comparison.}
\label{condprobplot}
\end{figure}

First we test the scaling behavior of the interaction, equation (\ref{scalelaw}).
In figure \ref{int} we have plotted the ratio $\overline{\textrm{I}} / \overline{\delta E}$
(see equations (\ref{avI}) and(\ref{levelvar})) for the three partitions and $1 / \sqrt{N}$ for
comparison.
The result confirms equation (\ref{scalelaw}). The values are slightly smaller than $1/\sqrt{N}$, which
is due to the neglect of the local terms of the Hamiltonian.

Then we calculate $w_j(\mu) \, \eta ( E_{\mu} - E_{j} )$ for all three partitions.
For each $j$ we have plotted the distribution $w_j(\mu) \, \eta ( E_{\mu} - E_{j} )$ versus
$x = ( E_{\mu} - E_{j} ) / \sqrt{\bra j \ve I^2 \ve j \ket}$ thus rescaling its width to unity.
Figure \ref{condprobplot} shows a superposition of all theese plots for one realisation.
For comparison we have plotted in the range,
where $\ve E_{\mu} - E_{j} \ve > \sqrt{\bra j \ve I^2 \ve j \ket}$, the functions 
$0.25 \, \exp ( - \beta \, \ve E_{\mu} - E_{j} \ve ) = 0.25 \, \exp ( - 0.5 \, \ve x \ve )$, taking
$\beta \sqrt{\bra j \ve I^2 \ve j \ket} = 0.5 < 1$.
The normalisation $0.25$ was deliberately
chosen to show that the distribution $w_j(\mu) \, \eta ( E_{\mu} - E_{j} )$ indeed decays fast enough
in that range. All numerical tests we made showed such a behavior.
%
%
%
%

\begin{figure}
\psfrag{0.4}{\small $0.4$}
\psfrag{0.3}{\small $0.3$}
\psfrag{0.2}{\small $0.2$}
\psfrag{0.1}{\small $0.1$}
\psfrag{0.21}{\small \hspace{-0.15cm} $0.20$}
\psfrag{0.15}{\small \hspace{-0.15cm} $0.15$}
\psfrag{0.11}{\small \hspace{-0.15cm} $0.10$}
\psfrag{0.05}{\small \hspace{-0.15cm} $0.05$}
\psfrag{1}{\small \raisebox{-0.1cm}{1}}
\psfrag{2}{\small \raisebox{-0.1cm}{2}}
\psfrag{3}{\small \raisebox{-0.1cm}{3}}
\psfrag{4}{\small \raisebox{-0.1cm}{4}}
\psfrag{d}{$dist(\rho, \tilde \rho)$}
\psfrag{b}{$\beta \, \lambda$}
\psfrag{n}{\raisebox{-0.1cm}{$N$}}
\psfrag{0.95}{\small \raisebox{0.82cm}{\hspace{-0.2cm} 1.00} \hspace{-0.8cm} 0.95}
\psfrag{0.91}{\small \hspace{-0.2cm} 0.90}
\psfrag{0.85}{\small \hspace{-0.2cm} 0.85}
\psfrag{0.81}{\small \hspace{-0.2cm} 0.80}
\psfrag{1.1}{\small \raisebox{0.4cm}{1}}
\psfrag{2.1}{\small \raisebox{0.4cm}{2}}
\psfrag{3.1}{\small \raisebox{0.4cm}{3}}
\psfrag{4.1}{\small \raisebox{0.4cm}{4}}
\psfrag{0.1}{\small 0.1}
\psfrag{0.2}{\small 0.2}
\psfrag{0.3}{\small 0.3}
\psfrag{0.4}{\small 0.4}
\psfrag{b}{\small $\beta \, \lambda$}
\psfrag{r}{\small \raisebox{0.3cm}{\hspace{-0.5cm} $\beta_{spec}/ \beta$}}
\psfrag{n}{\small $N$}
\twofigures[scale=0.65]{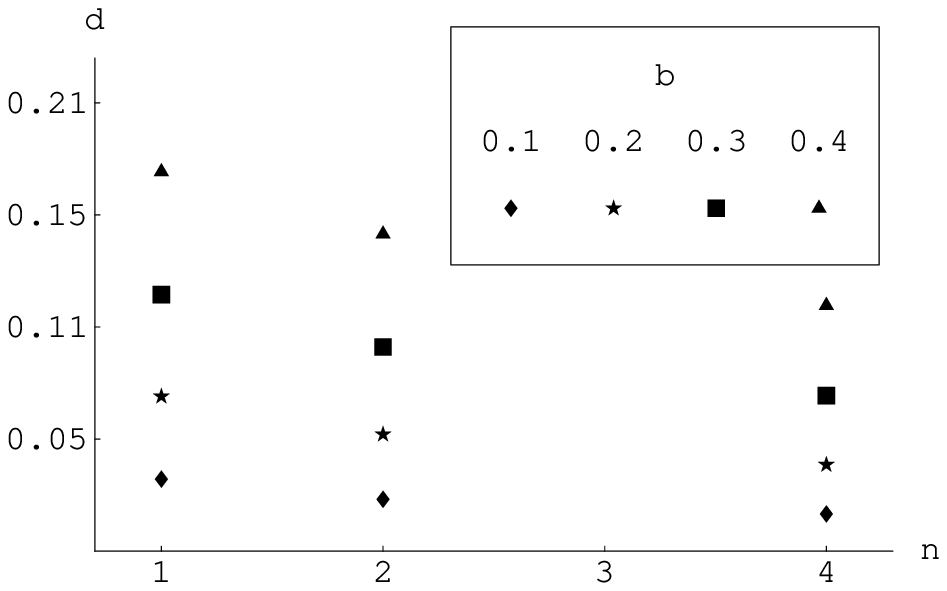}{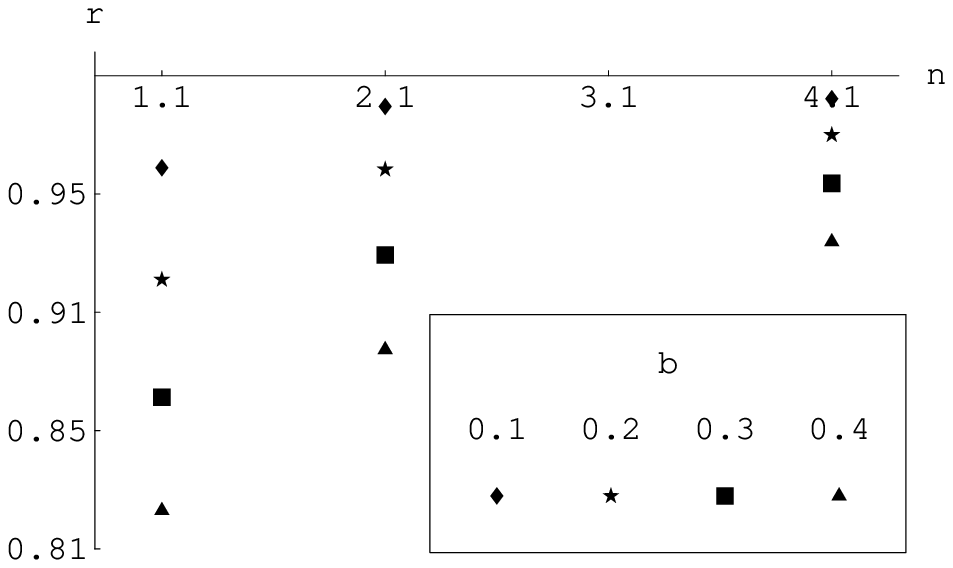}
\caption{Distance between the density matrix $\rho$
and a product of canonical density matrices of the subgroups
corresponding to the selected partition $\tilde \rho$. $N$ is the number of subsystems per group.}
\label{dist}
\caption{Ratios of the spectral inverse temperatures $\beta_{spec}$ of the subgroups to
the given inverse temperature $\beta$ of the total system. $N$ is the number of subsystems per group.}
\label{temp}
\end{figure}

Let us now assume the whole system to be in a thermal state with a reciprocal temperature $\beta$ and a 
density matrix according to equation (\ref{candens}).
We calculate the distance between this density matrix
and a product of canonical density matrices of the subgroups corresponding to a partition
\begin{equation}
\tilde \rho = \prod_{j=1}^{n_G} \otimes \: \rho_G^{can} (j)
\end{equation}
where $\rho_G^{can} (j)$ is of the same form as in equation (\ref{candens})
but only for one subgroup. As a measure for the distance we use
\begin{equation}
dist(\rho, \tilde \rho) = \sqrt{ \textrm{Tr} \left[ (\rho - \tilde \rho)^2 \right]}
\end{equation}
The result is shown in figure \ref{dist}:
The distance between the state of a global and a local
thermodynamical description is found to be approximately proportional to $\beta \: \lambda$ and decreases
as the group size $N$ increases. The points with $\sqrt{N} \, / \, (\beta \lambda) > 5$,
for which our estimates should apply, are below $0.1$, which supports equation (\ref{scalecond}).

Finally, to further confirm our findings, we calculate a ''spectral temperature'' of each subgroup
in the following way \cite{GemmerPhD}: 
The canonical density matrix of the whole system is transformed into the product basis
(\ref{prodstate}).
To each pair of states formed by an excited energy level $E_i$ and the ground level $E_0$ of a subgroup,
one can then assign a Boltzmann factor with an inverse temperature $\beta'$.
The spectral temperature is the sum of all these $\beta'$ weighted by the occupation probability
$p_i$ of the excited level $E_i$:
\begin{equation}
\label{specdef}
\beta_{spec} \equiv  - \sum_{i > 0}
\frac{p_i}{1 - p_0} \: \frac{\ln (p_{i}) - \ln (p_{0})}{E_{i} - E_0}
\end{equation}
Such a $\beta_{spec}$ can be defined for any state and coincides with the thermodynamical $\beta$
for a canonical state.
Since periodic boundary conditions were assumed, all subgroups of the same size $N$ have the same
temperature.
In figure \ref{temp} the ratios between $\beta_{spec}$
defined by equation (\ref{specdef}) and the inverse temperature of the total system $\beta$
are plotted versus group size $N$.
%
%
%
%
Here the points, which fulfill our criteria, lie above $0.9$, which again confirms our results.
The spectral temperatures of the subgroups are, in general, higher than the temperature of the total
system. This indicates that the state of a subgroup is more mixed than the state of the whole system,
which may be due to the entanglement of the subgroups caused by their mutual interaction \cite{Wang2002}.
%

\section{Conclusion}

We have studied a modular system of fixed size
composed of elementary subsytems with a finite energy spectrum
and nearest neighbour interactions. We have shown that if one forms groups of $N$ subsystems each,
the interaction between neighbouring groups scales as $N^{- \, 1/2}$.
We have then considered a chain of such interacting subsystems in a global thermal state with
canonical density matrix.
We have argued that due to the scaling property of the interaction, the reduced density matrix of each group
may be approximated by a canonical one with the same temperature as that of the total density matrix, if
$\sqrt{N} \gg \beta \: \overline{\delta E}$: The temperature
becomes an intensive quantity on a coarse-grained size and length-scale only.
In the same way, energy becomes more and more extensive
as the group-size increases.

We have tested this assertion numerically with a chain of 8 interacting spins and find that our predictions
are met in spite of the still small size of the total system.

Our studies
should be extended to the total system being in a local thermodynamical equilibrium only.
Here, heat conduction \cite{Cahill2003,Michel2003} becomes an interesting problem. For strong coupling,
meaningful temperature-profiles can be defined with limited resolution only
\cite{Hartmann2003}.

\acknowledgments
We thank M.\ Michel, H.\ Schmidt, M.\ Stollsteimer, and F. Tonner for
fruitful discussions.


\begin{thebibliography}{0}

\bibitem{Fisher1964}
  \Name{Fisher M.E.}
  \REVIEW{Arch. Ratl. Mech. Anal.}{17}{1964}{377}.

\bibitem{Ruelle1969}
  \Name{Ruelle D.}
  \Book{Statistical Mechanics}
  \Publ{W.A. Benjamin Inc., New York}
  \Year{1969}

\bibitem{Lebowitz1969}
  \Name{Lebowitz J.L., \and Lieb E.H.}
  \REVIEW{Phys. Rev. Lett.}{22}{1969}{631}.

\bibitem{GemmerOtte2001}
  \Name{Gemmer J., Otte A. \and Mahler G.}
  \REVIEW{Phys. Rev. Lett.}{86}{2001}{1927}.

\bibitem{Jensen1985}
  \Name{Jensen R. \and Shankar R.}
  \REVIEW{Phys. Rev. Lett.}{54}{1985}{1879}.

\bibitem{Cahill2003}
  \Name{Cahill D., Ford W., Goodson K., Mahan G., Majumdar A., Maris H., Merlin R. \and Phillpot S.}
  \REVIEW{J. Appl. Phys.}{93}{2003}{793}.

\bibitem{Gao2002}
  \Name{Gao Y. \and Bando Y.}
  \REVIEW{Nature}{415}{2002}{599}.

\bibitem{Pothier1997}
  \Name{Pothier H., Gu\'eron S., Brige N.O., Esteve D. \and Devoret M.H.}
  \REVIEW{Phys. Rev. Lett.}{79}{1997}{3490}.

\bibitem{Aumentado2001}
  \Name{Aumentado J., Eom J., Chandrasekhar V., Baldo P.M., \and Rehn L.E.}
  \REVIEW{Phys. Rev. Lett.}{88}{2002}{154301}.

\bibitem{Williams1986}
  \Name{Williams C.C. \and Wickramasinghe H.K.}
  \REVIEW{Appl. Phys. Lett.}{49}{1986}{1587}.

\bibitem{Varesi1998}
  \Name{Varesi J. \and Majumdar A.}
  \REVIEW{Appl. Phys. Lett.}{72}{1998}{37}.

\bibitem{Mahler1998}
  \Name{Mahler G.\and Weberruss V.}
  \Book{Quantum Networks}
  \Publ{Springer, Berlin}
  \Year{1998}

\bibitem{GemmerPhD}
  \Name{Gemmer J.}
  \Book{A Quantum Approach to Thermodynamics, PhD. Thesis}
  \Publ{University of Stuttgart, Stuttgart}
  \Year{2002}

\bibitem{Wang2002}
  \Name{Wang X.}
  \REVIEW{Phys. Rev. A}{66}{2002}{064304}.

\bibitem{Michel2003}
  \Name{Michel M., Hartmann M., Gemmer J. \and Mahler G.}
  \REVIEW{Europhys. J. B}{34}{2003}{325}.

\bibitem{Hartmann2003}
  \Name{Hartmann M., Mahler G. \and Hess O.}
  \REVIEW{}{}{}{in preparation}.

\end{thebibliography}
\end{document}